# Future Progress in Artificial Intelligence: A Survey of Expert Opinion

*Vincent C. Müller [a,b] & Nick Bostrom [a]*

[a] Future of Humanity Institute, Department of Philosophy & Oxford Martin School, University of Oxford. [b] Anatolia College/ACT, Thessaloniki

*Abstract:* There is, in some quarters, concern about high–level machine intelligence and superintelligent AI coming up in a few decades, bringing with it significant risks for humanity. In other quarters, these issues are ignored or considered science fiction. We wanted to clarify what the distribution of opinions actually is, what probability the best experts currently assign to high–level machine intelligence coming up within a particular time–frame, which risks they see with that development, and how fast they see these developing. We thus designed a brief questionnaire and distributed it to four groups of experts in 2012/2013. The median estimate of respondents was for a one in two chance that high-level machine intelligence will be developed around 2040-2050, rising to a nine in ten chance by 2075. Experts expect that systems will move on to superintelligence in less than 30 years thereafter. They estimate the chance is about one in three that this development turns out to be 'bad' or 'extremely bad' for humanity.

1. **Introduction**

Artificial Intelligence began with the "… conjecture that every aspect of learning or any other feature of intelligence can in principle be so precisely described that a machine can be made to simulate it." (McCarthy, Minsky, Rochester, & Shannon, 1955, p. 1) and moved swiftly from this vision to grand promises for general human-level AI within a few decades. This vision of general AI has now become merely a long-term guiding idea for most current AI research, which focuses on specific scientific and engineering problems and maintains a distance to the cognitive sciences. A small minority believe the moment has come to pursue general AI directly as a technical aim with the traditional methods − these typically use the label 'artificial general intelligence' (AGI) (see Adams et al., 2012).

If general AI were to be achieved, this might also lead to superintelligence: "We can tentatively define a superintelligence as *any intellect that greatly exceeds the cognitive performance of humans in virtually all domains of interest*." (Bostrom, 2014 ch. 2). One idea how superintelligence might come about is that if we humans could create artificial general intelligent ability at a roughly human level, then this creation could, in turn, create yet higher intelligence, which could, in turn, create yet higher intelligence, and so on … So we might generate a growth well beyond human ability and perhaps even an accelerating rate of growth: an 'intelligence explosion'. Two main questions about this development are when to expect it, if at all (see Bostrom, 2006; Hubert L. Dreyfus, 2012; Kurzweil, 2005) and what the impact of it would be, in particular which risks it might entail, possibly up to a level of existential risk for humanity (see Bostrom, 2013; Müller, 2014a). As Hawking et al. say "Success in creating AI would be the biggest event in human history. Unfortunately, it might also be the last, unless we learn how to avoid the risks." (Hawking, Russell, Tegmark, & Wilczek, 2014; cf. Price, 2013).

So, we decided to ask the experts what they predict the future holds − knowing that predictions on the future of AI are often not too accurate (see Armstrong, Sotala, & O Heigeartaigh, 2014) and tend to cluster around 'in 25 years or so', no matter at what point in time one asks.[1]

## 2. Questionnaire

### 2.1. Respondents

The questionnaire was carried out online by invitation to particular individuals from four different groups for a total of ca. 550 participants (see *Appendix 2*). Each of the participants got an email with a unique link to our site to fill in an online form (see *Appendix 1*). If they did not respond within 10 days, a reminder was sent, and another 10 days later, with the note that this is the last reminder. In the case of EETN (see below) we could not obtain the individual email addresses and thus sent the request and reminders to the members' mailing list. Responses were made on a single web page with one 'submit' button that only allowed submissions through these unique links, thus making non–invited responses extremely unlikely. The groups we asked were:

1. **PT–AI:** Participants of the conference on "Philosophy and Theory of AI", Thessaloniki October 2011, organized by one of us (see Müller, 2012, 2013).

---

[1] There is a collection of predictions on http://www.neweuropeancentury.org/SIAI-FHI_AI_predictions.xls

[2] A further, more informal, survey was conducted in August 2007 by Bruce J Klein (then of Novamente and the Singularity Institute) "… on the time–frame for when we may see greater–than–human level AI", with a few numerical results and interesting comments, archived on

   Participants were asked in November 2012, i.e. over a year after the event. The total of 88 participants include a workshop on "The Web and Philosophy" (ca. 15 people), from which a number of non–respondents came. A list of participants is on: http://www.pt–ai.org/2011/registered–participants
2. **AGI:** Participants of the conferences of "Artificial General Intelligence" (AGI 12) and "Impacts and Risks of Artificial General Intelligence" (AGI Impacts 2012), both Oxford December 2012. We organized AGI–Impacts (see Müller, 2014b) and hosted AGI 12. The poll was announced at the meeting of 111 participants (of which 7 only for AGI–Impacts) and carried out ca. 10 days later. The conference site is at: http://www.winterintelligence.org/oxford2012/
3. **EETN:** Members of the Greek Association for Artificial Intelligence (EETN), a professional organization of Greek published researchers in the field, in April 2013. Ca. 250 members. The request was sent to the mailing list. The site of EETN: http://www.eetn.gr/
4. **TOP100:** The 100 'Top authors in artificial intelligence' by 'citation' in 'all years' according to Microsoft Academic Search (http://academic.research.microsoft.com) in May 2013. We reduced the list to living authors, added as many as necessary to get back to 100, searched for professional e–mails on the web and sent notices to these.

The questionnaire was sent with our names on it and with an indication that we would use it for this paper and Nick Bostrom's new book on superintelligence (Bostrom, 2014) – our request email is in *Appendix 1*. Given that the respondent groups 1 and 2 attended conferences organized by us, they knew whom they were responding to. In groups 3 and 4 we would assume that the majority of experts would not know us, or even of us. These differences are reflected in the response rates.

These groups have different theoretical-ideological backgrounds: The participants of PT–AI are mostly theory–minded, mostly do not do technical work, and often have a critical view on large claims for easy progress in AI (Herbert Dreyfus was a keynote speaker in 2011). The participants of AGI are committed to the view that AI research should now return from technical details to 'artificial general intelligence' – thus the name AGI. The vast majority of AGI participants do technical work. The EETN is a professional association in Greece that accepts only published researchers from AI. The TOP100 group also works mostly in technical AI; its members are senior and older than the average academic; the USA is strongly represented.

Several individuals are members of more than one of these four sets and they were unlikely to respond to the same questionnaire more than once. So, in these cases, we sent the query only once, but counted a response for each set – i.e. we knew which individuals responded from the individual tokens they received (except in the case of EETN).

## 2.2. Response rates

1) PT–AI:     49%    43 out of 88
2) AGI:       65%    72 out of 111
3) EETN:      10%    26 out of 250
4) TOP100:    29%    29 out of 100
Total:        31%    170 out of 549

## 2.3. Methodology

In this field, it is hard to ask questions that do not require lengthy explanations or generate resistance in certain groups of potential respondents (and thus biased results). It is not clear what constitutes 'intelligence' or 'progress' and whether intelligence can be measured or at least compared as 'more' or 'less' as a single dimension. Furthermore, for our purposes we need a notion of intelligence at a level that may surpass humans or where technical intelligent systems might contribute significantly to research – but 'human–level intelligence' is a rather elusive notion that generates resistance. Finally, we need to avoid using terms that are already in circulation and would thus associate the questionnaire with certain groups or opinions, like "artificial intelligence", "singularity", "artificial general intelligence" or "cognitive system".

For these reasons, we settled for a definition that a) is based on behavioral ability, b) avoids the notion of a general 'human–level' and c) uses a newly coined term. We put this definition in the preamble of the questionnaire: "Define a *'high–level machine intelligence'* (HLMI) as one that can carry out most human professions at least as well as a typical human." (We still had one expert writing back to us that they could not say what a 'typical human' is – though they could be convinced to respond, after all.) In hindsight, it may have been preferable to specify what we mean by 'most' and whether we think of 'most professions' or of 'the professions most working people do'. One merit of our behavioral question is that having HLMI in our sense very likely implies being able to pass a classic Turing test.

To achieve a high response rate, we tried to have few questions with simple choices and eventually settled for four questions, plus three on the respondents. We tried to choose questions that would allow us to compare our results with those of earlier questionnaires – see below.

In order to improve on the quality of predictions, we tried to 'prime' respondents into thinking about what is involved in reaching HLMI before asking *when* they expect this. We also wanted to see whether people with a preference for particular approaches to HLMI would have particular responses to our central questions on prediction (e.g. whether people who think that 'embodied systems' are crucial expect longer than average time to HLMI). For these two purposes, we inserted a first question about contributing research approaches with a list to choose from – the options that were given are an eclectic mix drawn from many sources, but the particular options are not of much significance.

### 2.4. Prior work

A few groups have recently made attempts to gauge opinions. We tried to phrase our questions such that the answers can be compared to these earlier questionnaires. Notable are:

1. (Michie, 1973, p. 511f): "an opinion poll taken last year among sixty-seven British and American computer scientists working in, or close to, the machine intelligence field".
2. Questions asked live during the 2006 *AI@50* conference at Dartmouth College through a wireless voting device (VCM participated (see Müller, 2007)). Despite a short report on the conference in (Moor, 2006), the results were not published, but thankfully we were able to acquire them from the organizers James H. Moor and Carey E. Heckman – we publish a selection below.
3. (Baum, Goertzel, & Goertzel, 2011): participants of AGI 2009, not anonymous, on paper, 21 respondents, response rate unknown.[2]
4. (Sandberg & Bostrom, 2011): participants of *Winter Intelligence Conference 2011*, anonymous, on paper, 35 respondents, 41% response rate.

1. The reference by the famous AI researcher Donald Michie is very brief (all the details he gives are in the above quote) but of great of historical interest: 1972/3 were turning years for AI with the publication of Hubert Dreyfus' "What computers can't do" (Hubert L. Dreyfus, 1972), the "Lighthill Debates" on BBC TV (with Michie, McCarthy and R. Gregory) and the influential "Lighthill Report" (Lighthill, 1973). Michie's poll asked for the estimated number of years before "computing exhibiting intelligence at adult human level" and Michie's graph shows 5 data points:

---

[2] A further, more informal, survey was conducted in August 2007 by Bruce J Klein (then of Novamente and the Singularity Institute) "… on the time–frame for when we may see greater–than–human level AI", with a few numerical results and interesting comments, archived on https://web.archive.org/web/20110226225452/http://www.novamente.net/bruce/?p=54.

| Years | Percentage |
|---|---|
| 5 | 0% |
| 10 | 1% |
| 20 | 17% |
| 50 | 19% |
| >50 | 25% |

He also asked about "significant industrial spin-off", "contributions to brain studies" and "contributions from brain studies to machine intelligence". Michie adds "Of those responding to a question on the risk of ultimate 'takeover' of human affairs by intelligent machines, about half regarded it as 'negligible', and most of the remainder as 'substantial', with a view voting for 'overwhelming'." (Michie, 1973, p. 512).

2. AI@50 hosted many prominent AI researchers, including all living participants of the 1956 Dartmouth Conference, a set of DARPA-funded graduate students, plus a few theoreticians. The participants were asked 12 multiple choice questions on day one, 17 on day two and another 10 on day three. We select three results from day one here:

| 3.) The earliest that machines will be able to simulate learning and every other aspect of human intelligence: | | |
|---|---|---|
| Within 10 years | 6 | 5% |
| Between 11 and 25 years | 3 | 2% |
| Between 26 and 50 years | 14 | 11% |
| More than 50 years | 50 | 41% |
| Never | 50 | 41% |
| Totals | 123 | 100% |
| 5.) The earliest we will understand the basic operations (mental steps) of the human brain sufficiently to create machine simulation of human thought is: | | |
| today (we already understand enough) | 5 | 6% |
| within the next 10 years | 11 | 12% |
| within the next 25 years | 9 | 10% |
| within the next 50 years | 19 | 21% |
| within the next 100 years or more | 26 | 29% |
| never (we will never understand enough) | 19 | 21% |
| Totals | 89 | 100% |
| 6.) The earliest we will understand the architecture of the brain (how its organizational control is structured) sufficiently to create machine simulation of human thought is: | | |
| Within 10 years | 12 | 11% |

| Between 11 and 25 years | 15 | 14% |
| --- | --- | --- |
| Between 26 and 50 years | 24 | 22% |
| More than 50 years | 44 | 40% |
| Never | 15 | 14% |
| Totals | 110 | 100% |

3. Baum et al. asked for the ability to pass a Turing test, a third grade school year exam [i.e. for 9 year olds] and do Nobel Prize level research. They assume that all and only the intelligent behavior of humans is captured in the Turing test. The results they got for the 50% probability point were: 2040 (Turing test), 2030 (third grade), and 2045 (Nobel).

4. Sandberg and Bostrom's first question was quite similar to our 2nd (see below): "Assuming no global catastrophe halts progress, by what year would you assign a 10%/50%/90% chance of the development of human–level machine intelligence?" The median estimate of when there will be 50% chance of human–level machine intelligence was 2050. So, despite significant overlap with AGI 2009, the group asked by Sandberg and Bostrom in 2011 was a bit more guarded in their expectations.

We think it is worthwhile to make a new attempt because the prior ones asked specific groups and small samples, sometimes have methodological problems, and we also want to see how the answers change over time, or do not change – which is why tried to use similar questions. As explained below, we also think it might be worthwhile to repeat our questionnaire at a later stage, to compare results.

### 3.  Questions & Responses

#### 3.1. Research Approaches

"1. In your opinion, what are the research approaches that might contribute the most to the development of such HLMI?" [Selection from list, more than one selection possible.]

- Algorithmic complexity theory
- Algorithms revealed by computational neuroscience
- Artificial neural networks
- Bayesian nets
- Cognitive science
- Embodied systems
- Evolutionary algorithms or systems
- Faster computing hardware
- Integrated cognitive architectures

- Large–scale datasets
- Logic–based systems
- Robotics
- Swarm intelligence
- Whole brain emulation
- Other method(s) currently known to at least one investigator
- Other method(s) currently completely unknown
- No method will ever contribute to this aim

| | |
|---|---|
| Cognitive science | 47.9% |
| Integrated cognitive architectures | 42.0% |
| Algorithms revealed by computational neuroscience | 42.0% |
| Artificial neural networks | 39.6% |
| Faster computing hardware | 37.3% |
| Large-scale datasets | 35.5% |
| Embodied systems | 34.9% |
| Other method(s) currently completely unknown | 32.5% |
| Whole brain emulation | 29.0% |
| Evolutionary algorithms or systems | 29.0% |
| Other method(s) currently known to at least one investigator | 23.7% |
| Logic-based systems | 21.3% |
| Algorithmic complexity theory | 20.7% |
| No method will ever contribute to this aim | 17.8% |
| Swarm intelligence | 13.6% |
| Robotics | 4.1% |
| Bayesian nets | 2.6% |

The percentages here are over the total of responses. There were no significant differences between groups here, except that 'Whole brain emulation' got 0% in TOP100, but 46% in AGI. We did also not find relevant correlations between the answers given here and the predictions made in the following questions (of the sort that, for example, people who think 'embodied systems' crucial would predict later onset of HLMI).

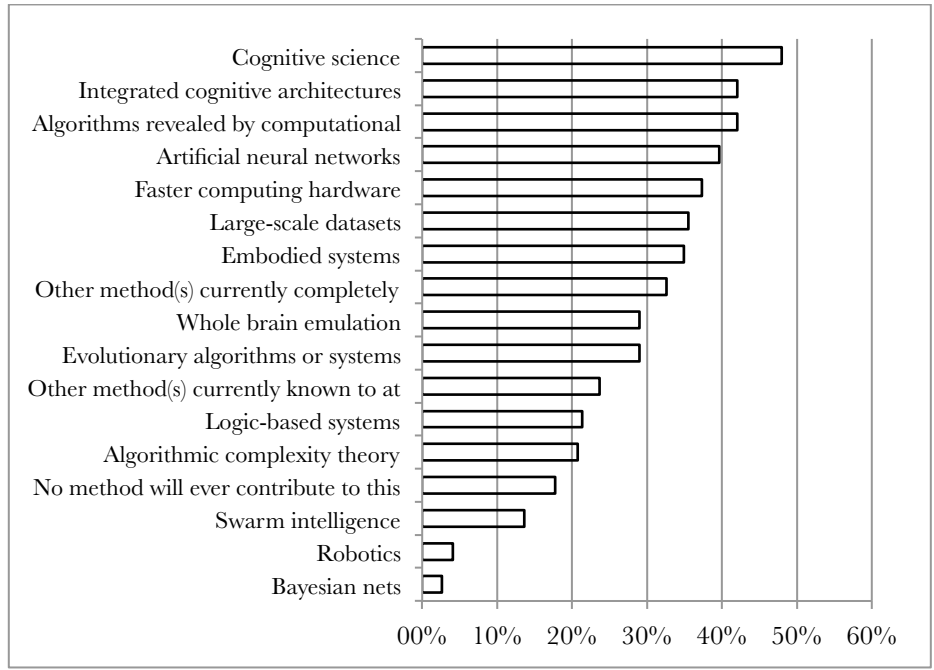

### 3.2. When HLMI?

"2. For the purposes of this question, assume that human scientific activity continues without major negative disruption. By what year would you see a (10% / 50% / 90%) probability for such HLMI to exist?" – For each of these three probabilities, the respondents were asked to select a year [2012–5000, in one-year increments] or check a box marked 'never'.

Results sorted by groups of respondents:

| PT-AI | Median | Mean | St. Dev. |
|---|---|---|---|
| **10%** | 2023 | 2043 | 81 |
| **50%** | 2048 | 2092 | 166 |
| **90%** | 2080 | 2247 | 515 |
| **AGI** | **Median** | **Mean** | **St. Dev.** |
| **10%** | 2022 | 2033 | 60 |
| **50%** | 2040 | 2073 | 144 |
| **90%** | 2065 | 2130 | 202 |
| **EETN** | **Median** | **Mean** | **St. Dev.** |
| **10%** | 2020 | 2033 | 29 |
| **50%** | 2050 | 2097 | 200 |
| **90%** | 2093 | 2292 | 675 |
| **TOP100** | **Median** | **Mean** | **St. Dev.** |
| **10%** | 2024 | 2034 | 33 |
| **50%** | 2050 | 2072 | 110 |
| **90%:** | 2070 | 2168 | 342 |
| **ALL** | **Median** | **Mean** | **St. Dev.** |
| **10%:** | 2022 | 2036 | 59 |
| **50%:** | 2040 | 2081 | 153 |
| **90%:** | 2075 | 2183 | 396 |

Results sorted by percentage steps:

| **10%** | **Median** | **Mean** | **St. Dev.** |
|---|---|---|---|
| **PT-AI** | 2023 | 2043 | 81 |
| **AGI** | 2022 | 2033 | 60 |
| **EETN** | 2020 | 2033 | 29 |
| **TOP100** | 2024 | 2034 | 33 |
| **ALL** | 2022 | 2036 | 59 |
| **50%** | **Median** | **Mean** | **St. Dev.** |
| **PT-AI** | 2048 | 2092 | 166 |
| **AGI** | 2040 | 2073 | 144 |
| **EETN** | 2050 | 2097 | 200 |
| **TOP100** | 2050 | 2072 | 110 |
| **ALL** | 2040 | 2081 | 153 |
| **90%** | **Median** | **Mean** | **St. Dev.** |
| **PT-AI** | 2080 | 2247 | 515 |
| **AGI** | 2065 | 2130 | 202 |
| **EETN** | 2093 | 2292 | 675 |
| **TOP100** | 2070 | 2168 | 342 |
| **ALL** | 2075 | 2183 | 396 |

Clicks of the 'never' box. These answers did *not* enter in to the averages above.

| Never | no. | % |
|---|---|---|
| **10%** | 2 | 1.2 |
| **50%** | 7 | 4.1 |
| **90%** | 28 | 16.5 |

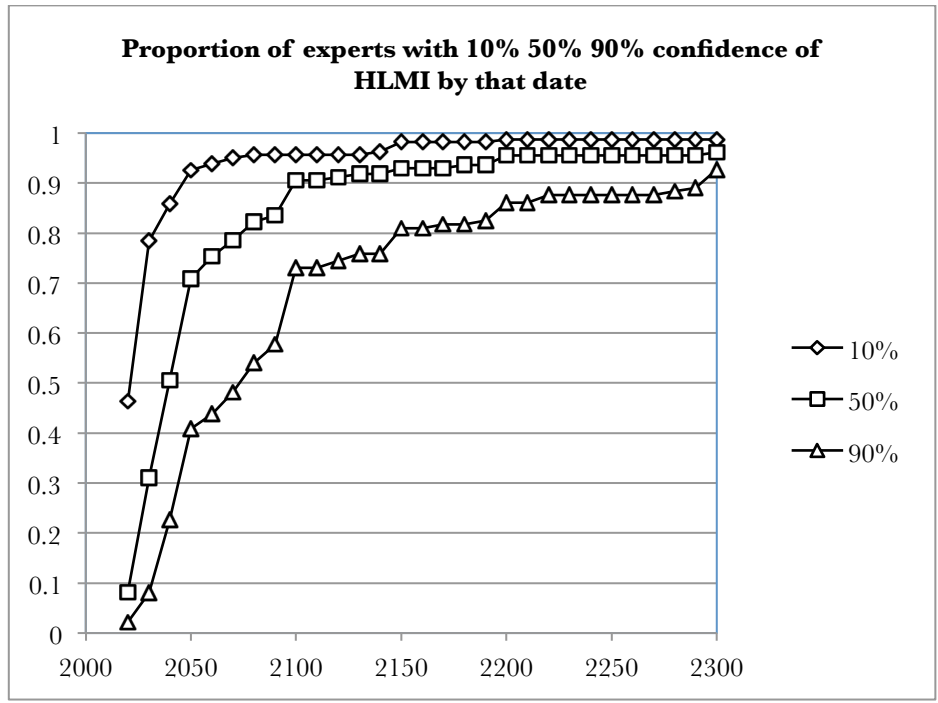

For the 50% mark, the overall median is 2040 (i.e. half of the respondents gave a year earlier than 2040 and half gave a year later than 2040) but the overall mean (average) is 2081. The median is always lower than the mean here because there cannot be outliers towards 'earlier' but there are outliers towards 'later' (the maximum possible selection was 5000, then 'never').

### 3.3. From HLMI to superintelligence

"3. Assume for the purpose of this question that such HLMI will at some point exist. How likely do you then think it is that within (2 years / 30 years) thereafter there will be machine intelligence that greatly surpasses the performance of every human in most professions?" – Respondents were asked to select a probability from a drop-down menu in 1% increments, starting with 0%.

For all respondents:

|  | Median | Mean | St. Dev. |
|---|---|---|---|
| **Within 2 years** | 10% | 19% | 24 |
| **Within 30 years** | 75% | 62% | 35 |

Median estimates on probability of superintelligence given HLMI in different groups of respondents:

|  | **2 years** | **30 years** |
|---|---|---|
| **PT-AI** | 10% | 60% |
| **AGI** | 15% | 90% |
| **EETN** | 5% | 55% |
| **TOP100** | 5% | 50% |

Experts allocate a low probability for a fast takeoff, but a significant probability for superintelligence within 30 years after HLMI.

### 3.4. The impact of superintelligence

"4. Assume for the purpose of this question that such HLMI will at some point exist. How positive or negative would be overall impact on humanity, in the long run? Please indicate a probability for each option. (The sum should be equal to 100%.)" – Respondents had to select a probability for each option (in 1% increments). The addition of the selection was displayed; in green if the sum was 100%, otherwise in red. The five options were: "Extremely good – On balance good – More or less neutral – On balance bad – Extremely bad (existential catastrophe)".

| % | **PT-AI** | **AGI** | **EETN** | **TOP100** | **ALL** |
|---|---|---|---|---|---|
| **Extremely good** | 17 | 28 | 31 | 20 | 24 |
| **On balance good** | 24 | 25 | 30 | 40 | 28 |
| **More or less neutral** | 23 | 12 | 20 | 19 | 17 |
| **On balance bad** | 17 | 12 | 13 | 13 | 13 |
| **Extremely bad (existential catastrophe)** | 18 | 24 | 6 | 8 | 18 |

Percentages here are means, not medians as in the other tables. There is a notable difference here between the 'theoretical' (PT-AI and AGI) and the 'technical' groups (EETN and TOP100).

### 3.5. Respondents Statistics

We then asked the respondents 3 questions about themselves:

1. "Concerning the above questions, how would you describe your own expertise?" (0 = none, 9 = expert)
   - Mean 5.85
2. "Concerning technical work in artificial intelligence, how would you describe your own expertise?" (0 = none, 9 = expert)
   - Mean 6.26
3. "What is your main home academic discipline?" (Select from list with 8 options: Biology/Physiology/Neurosciences – Computer Science – Engineering [non CS] – Mathematics/Physics – Philosophy – Psychology/Cognitive Science – Other academic discipline – None.) [Absolut numbers.]
   a. Biology/Physiology/Neurosciences   3
   b. Computer Science                 107
   c. Engineering (non CS)               6
   d. Mathematics/Physics               10
   e. Philosophy                        20
   f. Psychology/Cognitive Science      14
   g. Other academic discipline          9
   h. None                               1

And we finally invited participants to make a comment, plus a possibility to add their name, if they wished. (We cannot reproduce these here; but they are on our site, see below). A number of comments concerned the difficulty of formulating good questions, much fewer the difficulty of predicting.

## 4. Evaluation

### 4.1. Selection-bias in the respondents?

One concern with the selection of our respondents is that people who think HLMI is unlikely, or a confused idea, are less likely to respond (though we pleaded otherwise in the letter, see below). Here is a characteristic response from a keynote speaker at PT-AI 2011: "I wouldn't think of responding to such a biased questionnaire. … I think any discussion of imminent super–intelligence is misguided. It shows no understanding of the failure of all work in AI. Even just formulating such a questionnaire is biased and is a waste of time." (Hubert Dreyfus, quoted with permission). So, we tried to find out what the non-respondents think. To this end, we made a random selection of non-respondents from two groups (11 for PT-AI and 17 from TOP100) and pressured them via personal email to respond, explaining that this would help us understand bias. The two groups were selected because AGI appears already biased in the opposite direction and EETN appears very similar to TOP100 but for EETN we did not have the data to show us who responded and who did not. We got one additional response from PT-AI and two from TOP100 in this way.

For question 2 "… By what year would you see a (10% / 50% / 90%) probability for such HLMI to exist?" we compared the additional responses to the responses we already had from the same respective group (PT-AI and TOP100, respectively). We found the following differences:

|  | 10% | | 50% | | 90% | |
|---|---|---|---|---|---|---|
|  | mean | median | mean | median | mean | median |
| **PT-AI** | -12 | +8 | -9 | +55 | -2 | +169 |
| **TOP100** | -19 | -9 | -47 | -25 | -138 | -40 |

The one additional respondent from PT-AI expected HLMI earlier than the mean but later than the median, while the two respondents from TOP100 (last row) expected HLMI earlier than mean and median. The very small sample forbids confident judgment, but we found no support for the worry that the non-respondents would have been biased towards a later arrival of HLMI.

### 4.2. Lessons and outlook

We complement this paper with a small site on http://www.pt-ai.org/ai-polls/. On this site, we provide a) the raw data from our results [anonymous unless the participants decided to put their name on their responses], b) the basic results of the questionnaire, c) the comments made, and d) the questionnaire in an online format where anyone can fill it in. We expect that that online questionnaire will give us an interesting view of the 'popular' view of these matters and on how this view changes over time. In the medium run, it be interesting to do a longitudinal study that repeats this exact questionnaire.

We leave it to the reader to draw their own detailed conclusions from our results, perhaps after investigating the raw data. Let us stress, however, that the aim was to 'gauge the perception', not to get well-founded predictions. These results should be taken with some grains of salt, but we think it is fair to say that the results reveal a view among experts that AI systems will probably (over 50%) reach overall human ability by 2040-50, and very likely (with 90% probability) by 2075. From reaching human ability, it will move on to superintelligence in 2 years (10%) to 30 years (75%) thereafter. The experts say the probability is 31% that this development turns out to be 'bad' or 'extremely bad' for humanity.

So, the experts think that superintelligence is likely to come in a few decades and quite possibly bad for humanity – this should be reason enough to do research into the possible impact of superintelligence before it is too late. We could also put this more modestly and still come to an alarming conclusion: We know of no compelling reason to say that progress in AI will grind to a halt (though deep new insights might be needed) and we know of no compelling reason that superintelligent systems will be good for humanity. So, we should better investigate the future of superintelligence and the risks it poses for humanity.

## 5. Acknowledgements

Toby Ord and Anders Sandberg were helpful in the formulation of the questionnaire. The technical work on the website form, sending mails and reminders, database and initial data analysis was done by Ilias Nitsos (under the guidance of VCM). Theo Gantinas provided the emails of the TOP100. Stuart Armstrong made most graphs for presentation. The audience at the PT-AI 2013 conference in Oxford provided helpful feedback. Mark Bishop, Carl Shulman, Miles Brundage and Daniel Dewey made detailed comments on drafts. We are very grateful to all of them.

## 6. References

Adams, S., Arel, I., Bach, J., Coop, R., Furlan, R., Goertzel, B., . . . Sowa, J. F. (2012). Mapping the landscape of human-level artificial general intelligence. *AI Magazine, 33*(1), 25-42.

Armstrong, S., Sotala, K., & O Heigeartaigh, S. (2014). The errors, insights and lessons of famous AI predictions – and what they mean for the future. *Journal of Experimental and Theoretical Artificial Intelligence, 26*(3 - Special issue 'Risks of General Artificial Intelligence', ed. V. Müller), 317-342.

Baum, S. D., Goertzel, B., & Goertzel, T. G. (2011). How long until human-level AI? Results from an expert assessment. *Technological Forecasting & Social Change, 78*(1), 185-195.

Bostrom, N. (2006). How long before superintelligence? *Linguistic and Philosophical Investigations, 5*(1), 11-30.

Bostrom, N. (2013). Existential risk prevention as global priority. *Global Policy, 4*(1), 15-31.

Bostrom, N. (2014). *Superintelligence: Paths, dangers, strategies*. Oxford: Oxford University Press.

Dreyfus, H. L. (1972). *What computers still can't do: A critique of artificial reason* (2 ed.). Cambridge, Mass.: MIT Press.

Dreyfus, H. L. (2012). A history of first step fallacies. *Minds and Machines, 22*(2 - special issue "Philosophy of AI" ed. Vincent C. Müller), 87-99.

Hawking, S., Russell, S., Tegmark, M., & Wilczek, F. (2014). Transcendence looks at the implications of artificial intelligence - but are we taking AI seriously enough? *The Independent, 01.05.2014*.

Kurzweil, R. (2005). *The singularity is near: When humans transcend biology*. London: Viking.

Lighthill, J. (1973). Artificial intelligence: A general survey *Artificial intelligence: A paper symposion*. London: Science Research Council.

McCarthy, J., Minsky, M., Rochester, N., & Shannon, C. E. (1955). A proposal for the Dartmouth summer research project on artificial intelligence. Retrieved October 2006, from http://www-formal.stanford.edu/jmc/history/dartmouth/dartmouth.html

Michie, D. (1973). Machines and the theory of intelligence. *Nature, 241*(23.02.1973), 507-512.

Moor, J. H. (2006). The Dartmouth College artificial intelligence conference: The next fifty years. *AI Magazine, 27*(4), 87-91.

Müller, V. C. (2007). Is there a future for AI without representation? *Minds and Machines, 17*(1), 101-115.

Müller, V. C. (2014a). Editorial: Risks of general artificial intelligence. *Journal of Experimental and Theoretical Artificial Intelligence, 26*(3 - Special issue 'Risks of General Artificial Intelligence', ed. V. Müller), 1-5.

Müller, V. C. (Ed.). (2012). *Theory and philosophy of AI* (*Minds and Machines,* Vol. 22/2- Special volume): Springer.

Müller, V. C. (Ed.). (2013). *Theory and philosophy of artificial intelligence* (*SAPERE,* Vol. 5). Berlin: Springer.

Müller, V. C. (Ed.). (2014b). *Risks of artificial general intelligence* (*Journal of Experimental and Theoretical Artificial Intelligence,* Vol. (26/3) Special Issue): Taylor & Francis.

Price, H. (2013). Cambridge, Cabs and Copenhagen: My Route to Existential Risk. *The New York Times, 27.01.2013*. http://opinionator.blogs.nytimes.com/2013/01/27/cambridge-cabs-and-copenhagen-my-route-to-existential-risk/?_php=true&_type=blogs&_r=0

Sandberg, A., & Bostrom, N. (2011). Machine intelligence survey. *FHI Technial Report, 2011*(1). http://www.fhi.ox.ac.uk/research/publications/

## 7. Appendices

1. Questionnaire
2. Letter sent to participants

## **Appendix 1:** Online Questionnaire

Questionnaire: Future Progress in Artificial Intelligence | Phil...      http://www.pt-ai.org/questionnaire-future-progress-artificial-in...

### Questionnaire: Future Progress in Artificial Intelligence

(http://www.fhi.ox.ac.uk/) (http://www.futuretech.ox.ac.uk)

This brief questionnaire is directed towards researchers in artificial intelligence or the theory of artificial intelligence. It aims to gauge how people working in the field view progress towards its original goals of intelligent machines, and what impacts they would associate with reaching these goals.

Contribution to this questionnaire is by invitation only. If the questionnaire is filled in without such an invitation, the data will be disregarded.

Answers will be anonymized. Results will be made publicly available on the site of the Programme on the Impacts of Future Technology: http://www.futuretech.ox.ac.uk (http://www.futuretech.ox.ac.uk) .

Thank you for your time!

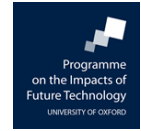
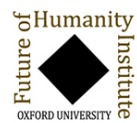

Vincent C. Müller (http://www.sophia.de) & Nick Bostrom (http://www.nickbostrom.com/)
University of Oxford
September 2012

#### A. The Future of AI

Define a **"high-level machine intelligence"** (HLMI) as one that can carry out most human professions at least as well as a typical human.

**1. In your opinion, what are the research approaches that might contribute the most to the development of such HLMI?:**

- ☐ Algorithmic complexity theory
- ☐ Algorithms revealed by computational neuroscience
- ☐ Artificial neural networks
- ☐ Bayesian nets
- ☐ Cognitive science
- ☐ Embodied systems
- ☐ Evolutionary algorithms or systems
- ☐ Faster computing hardware
- ☐ Integrated cognitive architectures
- ☐ Large-scale datasets
- ☐ Logic-based systems
- ☐ Robotics
- ☐ Swarm intelligence
- ☐ Whole brain emulation
- ☐ Other method(s) currently known to at least one investigator
- ☐ Other method(s) currently completely unknown
- ☐ No method will ever contribute to this aim

**2. Assume for the purpose of this question that human scientific activity continues without major negative disruption. By what year would you see a 10%/50%/90% probability for such HLMI to exist?**

|  | 10% | 50% | 90% |
|---|---|---|---|
| Year reached: | - | - | - |
| Never: | ☐ | ☐ | ☐ |

**3. Assume for the purpose of this question that such HLMI will at some point exist. How likely do you then think it is that within (2 years / 30 years) thereafter, there will be machine intelligence that greatly surpasses the performance of any human in most professions?**

|  | Within 2 years | Within 30 years |
|---|---|---|
| Probability: | - % | - % |

**4. Assume for the purpose of this question that such HLMI will at some point exist. How positive or negative would be the overall impact on humanity, in the long run? Please indicate a probability for each option. (The sum should be equal to 100%.)**

| Extremely good | On balance good | More or less neutral | On balance bad | Extremely bad (existential catastrophe) |
|---|---|---|---|---|
| - % | - % | - % | - % | - % |

Total: 0%

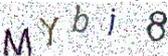

**Appendix 2:** Letter to participants (here TOP100)

Dear Professor [surname],

given your prominence in the field of artificial intelligence we invite you to express your views on the future of artificial intelligence in a brief questionnaire. The aim of this exercise is to gauge how the top 100 cited people working in the field view progress towards its original goals of intelligent machines, and what impacts they would associate with reaching these goals.

The questionnaire has 4 multiple choice questions, plus 3 statistical data points on the respondent and an optional 'comments' field. It will only take a few minutes to fill in.

Of course, this questionnaire will only reflect the actual views of researchers if we get nearly everybody to express their opinion. So, please do take a moment to respond, even (or especially) if you think this exercise is futile or misguided.

Answers will be anonymous. Results will be used for Nick Bostrom's forthcoming book "Superintelligence: Paths, Dangers, Strategies" (Oxford University Press, 2014) and made publicly available on the site of the Programme on the Impacts of Future Technology: http://www.futuretech.ox.ac.uk.

Please click here now:
[link]

Thank you for your time!

Nick Bostrom & Vincent C. Müller
University of Oxford